\documentclass[aps,prd,floatfix,superscriptaddress,preprintnumbers,showpacs,showkeys]{revtex4}
\usepackage{latexsym,epsfig,amsmath,amssymb}
\usepackage{amssymb}
\usepackage{amsmath}
\usepackage{amsfonts}
\usepackage{epsfig} 
\usepackage{verbatim} 
\usepackage{graphicx,color}

\newcommand{\seq}{\begin{subequations}}
\newcommand{\sen}{\end{subequations}}
\newcommand{\eq}{\begin{eqnarray}}
\newcommand{\en}{\end{eqnarray}}

\begin{document}

\title{Bounds on lepton flavor violating physics and decays of neutral mesons \\ 
from $\tau(\mu) \to 3 \ell,  \ell \gamma\gamma$-decays}

\noindent 
\author{Claudio O. Dib} 
\affiliation{Departamento de F\'\i sica y Centro Cient\'\i fico
Tecnol\'ogico de Valpara\'\i so-CCTVal,
Universidad T\'ecnica Federico Santa Mar\'\i a,
Casilla 110-V, Valpara\'\i so, Chile}
\author{Thomas Gutsche}
\affiliation{Institut f\"ur Theoretische Physik, Universit\"at T\"ubingen,
Kepler Center for Astro and Particle Physics,
Auf der Morgenstelle 14, D-72076, T\"ubingen, Germany}
\author{Sergey G. Kovalenko} 
\affiliation{Departamento de F\'\i sica y Centro Cient\'\i fico
Tecnol\'ogico de Valpara\'\i so-CCTVal,
Universidad T\'ecnica Federico Santa Mar\'\i a,
Casilla 110-V, Valpara\'\i so, Chile}
\author{Valery E. Lyubovitskij} 
\affiliation{Departamento de F\'\i sica y Centro Cient\'\i fico
Tecnol\'ogico de Valpara\'\i so-CCTVal,
Universidad T\'ecnica Federico Santa Mar\'\i a,
Casilla 110-V, Valpara\'\i so, Chile}
\affiliation{Institut f\"ur Theoretische Physik, Universit\"at T\"ubingen,
Kepler Center for Astro and Particle Physics,
Auf der Morgenstelle 14, D-72076, T\"ubingen, Germany}
\affiliation{Department of Physics, Tomsk State University,
634050 Tomsk, Russia}
\affiliation{Laboratory of Particle Physics, Tomsk Polytechnic University,
634050 Tomsk, Russia}
\author{Ivan Schmidt}
\affiliation{Departamento de F\'\i sica y Centro Cient\'\i fico
Tecnol\'ogico de Valpara\'\i so-CCTVal,
Universidad T\'ecnica Federico Santa Mar\'\i a,
Casilla 110-V, Valpara\'\i so, Chile}

\date{\today}

\begin{abstract} 

We study two- and three-body lepton flavor violating (LFV) decays 
involving leptons and neutral vector bosons 
$V=\rho^0, \omega, \phi, J/\psi, \Upsilon, Z^0$, 
as well as pseudoscalar $P=\pi^0, \eta, \eta', \eta_c$ and scalar 
$S=f_0(500), f_0(980), a_0(980), \chi_{c0}(1P)$ mesons,  
without referring to a specific mechanism of LFV realization.   
In particular, we relate the rates of the three-body LFV decays 
$\tau(\mu) \to 3 \ell$, where $\ell = \mu$ or $e$, to the two-body 
LFV decays $(V,P) \to \tau\mu(\tau e, \mu e)$, where $V$ and $P$ play 
the role of intermediate resonances in the decay process 
$\tau(\mu) \to 3 \ell$.  
From the experimental upper bounds for the branching ratios of 
$\tau(\mu) \to 3 \ell$ decays, we derive upper limits for the branching 
ratios of $(V,P) \to \tau\mu(\tau e, \mu e)$. 
We compare our results to the available experimental data and 
known theoretical upper limits from previous studies of LFV processes 
and find that some of our limits are several orders of magnitude more 
stringent. Using the idea of quark-hadron duality, we extract limits on 
various quark-lepton dimension-six LFV operators from data on  
lepton decays. Some of these limits are either new or stronger than those existing 
in the literature.

\end{abstract}

\pacs{11.30.Fs, 12.60.-i, 13.20.-v, 13.35.-r} 

\keywords{lepton flavor violation, leptons, 
vector, pseudoscalar and scalar mesons, $Z^0$ boson} 

\maketitle

\section{Introduction}
\label{sec:Introduction}

Search for lepton flavor violation (LFV) is an important probe of 
the possible physics beyond the Standard Model (SM).  
At present LFV is an established fact, since it has been already 
observed in neutrino oscillations, and therefore it is natural to expect 
that LFV is also going to manifest itself in the sector of charged leptons.  

A search strategy for LFV should consider those  processes 
which have the best prospect for discovery, both from the viewpoint of 
of their possible experimental identification 
and from theoretical limitations on the corresponding rates. 
The latter should incorporate the study of model independent 
relations between different processes, some of which are already strongly 
limited by experimental data.  

The three-body purely leptonic decays of $\mu$ and $\tau$ are among 
the most stringently constrained LFV processes, with the following current 
limits on their branching ratios~\cite{PDG} 
\begin{eqnarray}
\label{eq:ExpLim-mu}
&&{\rm Br}(\mu^{-} \to e^{-} e^{+} e^{-}) < 1.0 \times 10^{-12}\,,\\
\label{eq:ExpLim-tau-e}
&&{\rm Br}(\tau^{-} \to  e^{-} e^{+} e^{-}) < 2.7 \times 10^{-8}\,,\\
\label{eq:ExpLim-tau-mu}
&&{\rm Br}(\tau^{-} \to  \mu^{-} e^{+} e^{-}) < 1.8 \times 10^{-8}\,,\\
\label{eq:ExpLim-mu-e-2g}
&&{\rm Br}(\mu^{-} \to  e^{-} \gamma\gamma) \hspace{12pt} < 7.2 \times 10^{-11}\,.
\end{eqnarray}

\begin{figure}[htb]
\vspace*{-3cm}
\includegraphics[scale=0.45]{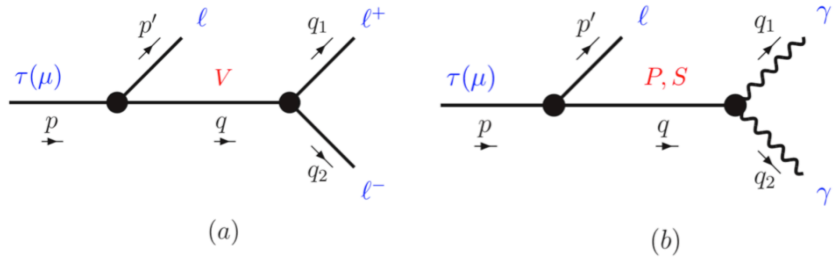}
\noindent
\mbox{}\\[-35mm]
\caption{Three-body LFV decays:  (a) $\tau(\mu) \to 3 \ell$ and 
(b) $\tau(\mu) \to \ell \, \gamma\gamma$.
\label{fig:tau3ell}}
\end{figure}
The purpose of the present paper is to relate the three-body lepton and 
lepton-photon decays of $\mu$ and $\tau$ (see Fig.~\ref{fig:tau3ell}) 
to the two-body LFV decays of neutral vector bosons and pseudoscalar mesons, 
and to give upper limits for these two-body branching ratios in a model 
independent way. 
We also study the LFV dimension-six quark-lepton effective 
operators underlying these processes and derive limits on their scales 
from the limits~(\ref{eq:ExpLim-mu})-(\ref{eq:ExpLim-tau-mu}). 
 
There already exist in the literature similar studies of limits on 
the two-body LFV decays of vector mesons/bosons  
$V = \rho^0, \omega, \phi, J/\psi, \Upsilon, Z^0 \to \mu^{\pm}e^{\mp}$, 
which use the constraint given in~(\ref{eq:ExpLim-mu})  and 
unitarity-inspired arguments~\cite{Nussinov:2000nm}.

The idea of using effective quark-lepton and
hadron-lepton Lagrangians for studying 
LFV processes (lepton-flavor changing decays, lepton-flavor
conversion, double beta decay) have been proposed and developed in
Refs.~\cite{Vergados:1985pq}-\cite{Kuno:1999jp}
and further used in a series of papers
(see, e.g., Refs.~\cite{Faessler:2004jt}-\cite{Davidson:2018kud}).
In particular, in Refs.~\cite{Faessler:1996ph}, the on-mass-shell
matching condition between the quark-level effective Lagrangian
and the effective hadronic-level (e.g., nucleon) Lagrangian was proposed, 
which sets the relations between the couplings at the quark level to those 
at the hadronic level.
In a series of papers~\cite{Faessler:2004jt}-\cite{Gonzalez:2013rea},  
 $\mu^--e^-$ conversion in nuclei was studied 
in the framework of an effective 
Lagrangian approach, without referring to any specific realization of 
the physics beyond the SM responsible for LFV. 
Limits on various LFV couplings of vector and scalar mesons 
to the $\mu-e$ current were derived from the existing experimental data on 
$\mu^{-}-e^{-}$ conversion in nuclei. 
Here, we extend the application of these techniques, in order to extract 
limits on two-body LFV decays of vector and pseudoscalar mesons by searching 
for LFV three-lepton decays of tau leptons and muons. 

The paper is organized as follows. 
In Sec.~II, we introduce the relevant effective quark-lepton and 
meson-lepton LFV operators, without referring to specific mechanisms of LFV. 
In Sec.~III, we derive the relations between three-body lepton LFV 
decays and two-body LFV meson decays, which is done by taking into 
account the contribution of neutral vector and pseudoscalar mesons in the 
three-body lepton LFV process. With these relations, we set the limits on 
the two-body LFV meson decays. In Sec.~IV, we derive the relations between 
branching ratios of two-body LFV decays of the same quark content and examine 
the limits on the effective quark-lepton operators from purely leptonic 
processes. Section V contains our summary and conclusions.  


\section{Effective Quark-Lepton and Meson-Lepton LFV operators}
\label{sec:Effective Lepto-quark and Lepto-Meson LFV operators}

Let us assume generic LFV sources, leading to $\tau\rightarrow \mu(e) e e$ 
and $\mu\rightarrow 3e$ decays, in the form of effective operators as the 
low-energy limit of a renormalizable ``fundamental'' LFV theory 
at a scale $\Lambda$. The leading-order operators have been 
proposed in Refs.~\cite{Vergados:1985pq}-\cite{Kuno:1999jp}.  
The set of these operators can be written as 
\begin{eqnarray}\label{eq:EffOperLept-1}
\mbox{4-lepton:}&&\mathcal{L}_{4\ell} 
= \frac{1}{\Lambda^2} \, 
\sum\limits_{(IJ)} \, 
C^{\Gamma_I\Gamma_J}_{\ell_{1}\ell_{2}} \, 
[\, \bar{\ell}_{1} \Gamma_I \ell_{2} \, ] \cdot 
[\,\bar{e}   \Gamma_J e\, ]  + {\rm H.c.} 
\\
\label{eq:MagneticOper}
\mbox{Magnetic:}&&\mathcal{L}_{M} = 
\frac{1}{\Lambda} 
\bigg( \, 
\bar{C}^{T}_{\ell_{1}\ell_{2}}  [\, \bar\ell_{1}  \sigma_{\mu\nu} \ell_{2}  \, ]  
+  
 \bar{C}^{T5}_{\ell_{1}\ell_{2}}  [\,  \bar\ell_{1}  
\sigma_{\mu\nu} \gamma_{5}\ell_{2}  \, ]  \,
 \bigg) \, F^{\mu\nu}\, + \, {\rm H.c.}\,,
\\
\label{eq:l-q-operators}
\mbox{Quark-Lepton:}&&\mathcal{L}_{\ell q} = 
\frac{1}{\Lambda^2} \, 
\sum\limits_{(IJ)} \, 
C^{\Gamma_I\Gamma_J}_{if,\ell_{1}\ell_{2}} \, 
 [\,  \bar{\ell}_{1} \Gamma_I \ell_{2}  \, ]  \cdot 
 [\,  \bar{q}_{f}    \Gamma_J q_{i}  \, ]  + {\rm H.c.} 
\nonumber\\
&=&\frac{1}{\Lambda^2} 
\, \bigg( 
\ 
C^{SS}_{if,\ell_{1}\ell_{2}} \, 
[\bar{\ell}_{1}  \ell_{2} ] \cdot  [ \bar{q}_{f}    q_{i}]  
\ +\
C^{PS}_{if,\ell_{1}\ell_{2}} \, 
[\bar{\ell}_{1}  \gamma^5 \ell_{2} ] \cdot  [ \bar{q}_{f}    q_{i}] 
\ +\
C^{SP}_{if,\ell_{1}\ell_{2}} \, 
 [\,  \bar{\ell}_{1} \ell_{2}  \, ]  \cdot  [\,   \bar{q}_{f} \gamma_5  q_{i}  \, ] 
\nonumber
\\
&& \hspace{-6pt}+\ 
 C^{PP}_{if,\ell_{1}\ell_{2}} \, 
 [\,  \bar{\ell}_{1} \gamma^5 \ell_{2}  \, ]  \cdot   [\, \bar{q}_{f} \gamma_5  q_{i}  \, ] 
\ +\
C^{VV}_{if,\ell_{1}\ell_{2}} \, 
 [\,  \bar{\ell}_{1} \gamma^\mu \ell_{2}  \, ] \cdot  
 [\,  \bar{q}_{f} \gamma_\mu q_{i}  \, ] 
\  +\ 
 C^{AV}_{if,\ell_{1}\ell_{2}} \, 
  [\,  \bar{\ell}_{1} \gamma^\mu \gamma^5 \ell_{2}  \, ] 
\cdot  
 [\,  \bar{q}_{f} \gamma_\mu q_{i}  \, ] 
\nonumber
\\
&& \hspace{-6pt}+\  
C^{VA}_{if,\ell_{1}\ell_{2}} \, 
 [\,  \bar{\ell}_{1} \gamma^\mu \ell_{2}  \, ] 
\cdot  
 [\,  \bar{q}_{f} \gamma_\mu\gamma_5 q_{i} 
 \, ] 
\  +\ 
 C^{AA}_{if,\ell_{1}\ell_{2}} \, 
  [\,  \bar{\ell}_{1} \gamma^\mu\gamma^5\, \ell_{2}  \, ]  \cdot  
 [\,  \bar{q}_{f} \gamma_\mu\gamma_5 q_{i}  \, ] 
\nonumber\\ 
&& \hspace{-6pt}+\  
C^{TT}_{if,\ell_{1}\ell_{2}} \, 
 [\,  \bar{\ell}_{1} \sigma^{\mu\nu} \ell_{2}  \, ]  \cdot 
 [\,  \bar{q}_{f} \sigma_{\mu\nu} q_{i}  \, ] 
\ \bigg)
 + {\rm H.c.} \,,
\label{eq:T-term-1}
\end{eqnarray} 
where $\ell= \mu, e$ and 
$F^{\mu\nu} = \partial^{\mu}A^{\nu}-\partial^{\nu}A^{\mu}$ 
is the electromagnetic field tensor. 
In (\ref{eq:EffOperLept-1}) and (\ref{eq:l-q-operators}),  
we use $I,J= S, P, V, A, T$ and
$\Gamma_{I,J} = 1, \gamma_{5}, \gamma_{\mu}, \gamma_{\mu}\gamma_{5}, 
\sigma_{\mu\nu}$, so that 
the summation runs over $(IJ) = (SS), (PS), (SP), (PP), 
(AV), (VV), (VA), (AA), (TT)$. In Eq.~(\ref{eq:T-term-1}),  
we displayed the terms in the sum explicitly.
After specifying all possible Lorentz structures 
in Eqs.~(\ref{eq:EffOperLept-1}) and 
(\ref{eq:l-q-operators}), we used the identity 
$
\bar{a} \sigma^{\mu\nu}\gamma_{5} b \cdot  
\bar{c} \sigma_{\mu\nu}\gamma_5 d =\bar{a} \sigma^{\mu\nu} b \cdot  
\bar{c} \sigma_{\mu\nu} d.
$
Here, we denoted the LFV scale by $\Lambda$.

\begin{center}
\begin{figure}[htb] 
\vspace{-2.6cm}
\includegraphics[scale=0.35,angle=0]{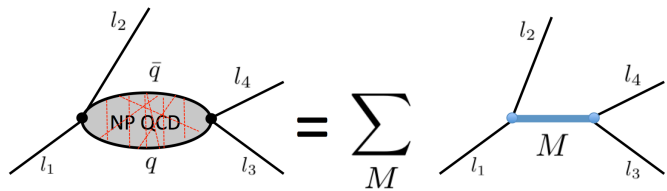}
\vspace{-2.4cm}
\caption{
Quark-lepton contact interaction contribution to 
$l_{1}\rightarrow l_{2} l_{3} l_{4}$ via meson exchange 
according to quark-hadron duality.}
\label{fig-duality}
\end{figure}
\end{center}

The operators~(\ref{eq:EffOperLept-1}) and~(\ref{eq:MagneticOper}) lead to 
tree-level contributions to $\tau\rightarrow \ell e e$, while 
the dipole-type  operator~(\ref{eq:MagneticOper}) directly contributes to 
$\tau\rightarrow \ell \gamma$. 
Limits on the scales of these operators are readily extracted from 
data~\cite{PDG} and can be found in the literature (see, for instance, 
Ref.~\cite{Lims-4-lept-M}). 
The quark-lepton LFV operators~(\ref{eq:l-q-operators}) 
have been studied by many authors, which consider the two-body decays
$\tau\rightarrow \ell M$, $M\rightarrow \ell_{1} \ell_{2}$, 
deep inelastic conversion 
$\tau(\mu) q \rightarrow \ell q$~\cite{Gninenko:2018num} 
as well as nuclear $\mu-e$- conversion (for a recent review see, for instance, 
Ref.~\cite{Nucl-MuE}). The existing data on the rates of these processes 
allowed extraction of rather stringent limits on the scale of 
the corresponding operators~(\ref{eq:l-q-operators}), which also   
contribute to leptonic LFV decays of 
mesons $M\rightarrow \ell_{1} \ell_{2}$ at tree level. 
At one-loop level they contribute 
to purely leptonic LFV processes $\tau^{-}\rightarrow \mu^{-} e^{+} e^{-}$ 
and $\mu^{-}\rightarrow e^{-} e^{+} e^{-}$. 
However, quark-hadron duality \cite{Nussinov:2000nm} relates
these loop contributions, taking into account nonperturbative QCD effects, 
with the sum over the tree-level contributions (Fig.~\ref{fig-duality}) of all 
the intermediate meson states with the allowed quantum numbers. 

Therefore, effectively the operators in Eq.~(\ref{eq:l-q-operators}) 
trigger tree-level contributions to 
$\ell_{1}\rightarrow \ell_{2} \, e^{+} e^{-}$ 
via intermediate meson states. The relevant meson-lepton vertices 
involving vector $V$, axial $A$, pseudoscalar $P$, and scalar $S$ mesons 
with quantum numbers $J^{PC} = 1^{--}$, $1^{++}$ ($1^{+-}$), $0^{-+}$, 
and $0^{++}$, respectively, are   
\begin{eqnarray}
\label{eq:Meson-Lepton-LFV-Vert}
\mathcal{L}_{\ell M} &=& 
V_{\mu} \left(g^{(V)}_{V\ell_{1}\ell_{2}} \, 
[\bar{\ell}_{1} \gamma^{\mu} \ell_{2}]+
g_{V\ell_{1}\ell_{2}}^{(A)} \, 
[\bar{\ell}_{1} 
\gamma^{\mu}\gamma_{5} \ell_{2} ] \right) +
A_{\mu} \left(g^{(V)}_{A\ell_{1}\ell_{2}} \, 
[ \bar{\ell}_{1} \gamma^{\mu} \ell_{2} ]+
g_{A\ell_{1}\ell_{2}}^{(A)} \,
[ \bar{\ell}_{1} 
\gamma^{\mu}\gamma_{5} \ell_{2} ] \right) 
\nonumber\\
&+&
\frac{g^{(T)}_{V\ell_1\ell_2}}{M_V} \, 
F^{V}_{\mu\nu}\, \left[\bar{\ell}_{1} \sigma^{\mu\nu} \ell_{2}\right]  
+\frac{g^{(T)}_{A\ell_1\ell_2}}{M_A} \, 
F^{A}_{\mu\nu}\, \left[\bar{\ell}_{1} \sigma^{\mu\nu} \gamma^5 \ell_{2}\right]
\nonumber\\
&+&S\, \left(g^{(S)}_{S\ell_{1}\ell_{2}} \,
[ \bar{\ell}_{1}  \ell_{2} ]
+ 
g_{S\ell_{1}\ell_{2}}^{(P)} \, 
[ \bar{\ell}_{1} \gamma_{5} \ell_{2}] \right) + 
P\, \left( i g^{(S)}_{P\ell_{1}\ell_{2}} \, [ \bar{\ell}_{1}   \ell_{2} ] + 
i g_{P\ell_{1}\ell_{2}}^{(P)} \, [\bar{\ell}_{1} \gamma_{5} \ell_{2} ]
\right) 
\nonumber\\
&+&
\frac{\partial_\mu P}{M_P} \, 
\left(g^{(V)}_{P\ell_1\ell_2} [ \bar{\ell}_{1} \gamma^\mu \ell_{2} ] +
g^{(A)}_{P\ell_1\ell_2}  [ \bar{\ell}_{1} \gamma^\mu\gamma^5 \ell_{2} ] \right) 
\,+\, {\rm H.c.} 
\end{eqnarray}
Here we introduced the notation 
$F^{M}_{\mu\nu} = \partial_{\mu}M_{\nu}- \partial_{\nu}M_{\mu}$ 
(with $M=V ,A$) for the field tensors of the vector and axial mesons, 
respectively.  
Obviously, the lightest mesons dominate in the diagram in 
Figs.~\ref{fig:tau3ell}(a) and 1(b), 
because the contributions of meson resonances to the three-body LFV 
decays scale as $1/M^4$, where $M$ is the mass of intermediate meson. 
     
In the next sections, we shall use the effective hadronic-level 
Lagrangian of Eq.~(\ref{eq:Meson-Lepton-LFV-Vert}) in order to constrain 
the quark-lepton operators of Eq.~(\ref{eq:l-q-operators}),  
using the bounds given in 
Eqs.~(\ref{eq:ExpLim-mu})-(\ref{eq:ExpLim-tau-mu}). This is done by
applying an appropriate matching condition at the hadronization scale. 
In this way, we shall constrain the vector and tensor operators related to
the corresponding vector boson contribution 
$(V = \rho^0, \omega, \phi, J/\psi, \Upsilon, Z^0)$ 
to the processes $\tau(\mu) \to 3 \ell$, and also constrain 
the pseudoscalar and scalar operators from the contribution of 
the pseudoscalar $(P = \pi^0, \eta, \eta', \eta_c(1S))$ and 
scalar meson states $(S = f_{0}(500), f_{0}(980), a_{0}(980), \chi_{c0}(1P))$ 
to the processes $\tau(\mu) \to \ell \gamma\gamma$. 
Expressions for the LFV two-body decay widths of different meson states 
are shown in Appendix~\ref{App:B}. 

Let us recall a key point of the present study: non-perturbative 
QCD effects leading to the formation of the $M$ meson bound states 
in the intermediate state of $\ell_{1}\rightarrow \ell_{2}\, e^{+}e^{-}$ 
are taken into account according
to the quark-hadron duality, via two parameters: the meson masses $M_{M}$ and 
their leptonic decay constants $f_{M}$.  
Numerical values of these parameters are known either from direct 
experimental measurements, from lattice simulations or some reliable 
models. The list of these parameters are given in Appendix \ref{App:A}. 
We shall study these meson exchange  mechanisms in the next sections.


\section{Relations between three- and two-body LFV decays}
\label{sec:Relations between Tree- and two-body LFV decays}

Here we derive unitarity-inspired relations between the three-body 
lepton decays and the two-body vector, scalar, and pseudoscalar 
meson decays. 
Unitarity implies the contribution of all intermediate meson states to 
$\tau,\mu\rightarrow\ell e e, \ell\,  \gamma \gamma$.  
Following Ref.~\cite{Nussinov:2000nm}, we retain as a good approximation  
only the lightest mesons,
so that their contributions are described by the meson exchange diagrams 
in Figs.~\ref{fig:tau3ell}, with the LFV vertices given 
by the Lagrangian~(\ref{eq:Meson-Lepton-LFV-Vert}). 
We shall not consider flavored mesons, because their decay 
rates \mbox{$M\rightarrow e^{+}e^{-}, \gamma\gamma$}, which enter 
in the above-mentioned relations, are GIM-suppressed, and this does 
not allow us to derive significant limits for their LFV decays.  

\subsection{Vector mesons}
\label{sec:Vector meson LFV decays}

Let us consider the vector mesons
$V = \rho^0, \omega, \phi, J/\psi, \Upsilon, Z^0$. 
Our goal is to analyze their contribution to $\mu,\tau\rightarrow 3\ell$ 
decays. For the case of 
$\mu^{-}\rightarrow e^{-} e^{+}e^{-}$ and vector 
mesons, this was done in Ref.~\cite{Nussinov:2000nm}. 

Neglecting the final lepton masses, for the muon decay rates we have
\begin{eqnarray}\label{eq:mu-3e-1} 
\Gamma(\mu^{-}\to e^{-} e^{+}e^{-}) &=& 
\kappa \frac{\tilde{g}_{V\mu e}^{2} 
\tilde{g}_{Vee}^{2}}{M_{V}^{4}}\\
\label{eq:mu-3e-2} 
\Gamma(\mu^{-}\to e^{-} \bar\nu_{e} \nu_{\mu}) &=& 
\Gamma (\mu \to All) = \kappa \frac{g_{W}^{4}}{M_{W}^{4}}\,, 
\end{eqnarray}
where $M_V$ is the vector meson/boson mass, 
$\kappa = M_\mu^5/(384 \pi^3)$ 
is a kinematic-spin factor common to all decay modes involving 
vector mesons in the intermediate state, while
$g_{W}$ and $M_{W}$ are the electroweak coupling and the $W$ boson mass, respectively 
(here $g_W$ is normalized so that the Fermi coupling is 
$G_F/\sqrt{2} =  g_W^2 /(2 M_W^2)$).  
By definition 
$\tilde{g}_{V\mu e}^{2} = |g^{(V)}_{V\mu e}|^{2}+|g^{(A)}_{V\mu e}|^{2}$. 
Then one finds for the LFV branching ratio 
\begin{eqnarray}\label{eq:BrMu3E} 
{\rm Br}(\mu^{-}\to e^{-} e^{+}e^{-}) &=& 
\frac{\tilde{g}_{V \mu e}^2 
\tilde{g}_{Vee}^2}{M_V^4} \ 
\frac{M_W^4}{g_W^4}\,.
\end{eqnarray}
Formulae for the meson two-body decay rates are given 
in Appendix~\ref{App:B}. Neglecting the 
final lepton masses they can be written as 
\begin{eqnarray}
\label{eq:V2LL-1}
& &\Gamma(V \to e^+e^-) = a\, 
\tilde{g}_{Vee}^2 \, M_V 
\,, \\ 
\label{eq:V2LL-2}
& &\Gamma(V \to \mu^\pm e^\mp) = a\, 
\tilde{g}_{V \mu e}^2 \, M_V 
\,, \\
\label{eq:V2LL-3}
& &\Gamma(W \to e\bar\nu_e) = a\,  g_W^2 \, M_W\, , \qquad
\end{eqnarray}
where $a = 1/(12\pi)$ is a kinematic factor common to all these processes. 
 
The branching ratio of Eq.~\eqref{eq:BrMu3E} can then be written  
in terms of the two-body decay rates as: 
\begin{eqnarray}\label{eq:NussinovFivRes-1}   
{\rm Br}(\mu \to 3 e) &=& 
\frac{\Gamma(V \to \mu  e) \, 
\Gamma(V \to e^+e^-)}{\Gamma(W \to e \bar\nu_e)^2} \, 
\biggl(\frac{M_W}{M_V}\biggr)^6 \,.
\end{eqnarray}

For the case of $\tau^{-}\rightarrow e^{-}(\mu^{-}) e^{+}e^{-}$ 
there are two main differences with respect to the muon decays: 
(i) due to the large mass of the $\tau$ lepton, 
there are some on-mass-shell meson contributions to this process; 
(ii) the $\tau$ decay width is not purely an electroweak quantity, i.e.
\mbox{$\Gamma(\tau \to All) \neq \Gamma (\tau \to \ell \bar\nu \nu)$}, 
since it contains hadronic channels. The latter suffer from considerable 
theoretical uncertainties. 
However, the tau decay width is an experimentally well measured observable~\cite{PDG}. 
Combining the above formulae~(\ref{eq:mu-3e-1}), (\ref{eq:mu-3e-2}) 
and~(\ref{eq:V2LL-1})-(\ref{eq:V2LL-3}) with the corresponding replacements, 
for $M_{V}> M_{\tau}$ we find: 
\begin{eqnarray}\label{eq:TauLEE-1}     
{\rm Br}(\tau^{-}\rightarrow \ell^{-} e^{+}e^{-}) &=&  
 \frac{\Gamma(V \to \tau \ell) \, 
\Gamma(V \to e^+e^-)}{\Gamma(W \to e \bar\nu_e)^2} \,                 
 \frac{\Gamma(\mu^{-}\to e^{-} \bar\nu_{e} \nu_{\mu})}{\Gamma(\tau \to All)}
  \left(\frac{M_W}{M_V}\right)^6   
\left(\frac{M_{\tau}}{M_{\mu}}\right)^5\,, 
\end{eqnarray}
and for $M_{V}<M_{\tau}$ 
\eq\label{eq:TauLEE-2} 
{\rm Br}(\tau^{-}\rightarrow \ell^{-} e^{+}e^{-}) &=& 
{\rm Br}(\tau \to V \ell)\,  {\rm Br}(V \to e^{+}e^{-})\,  
\en
where $\ell = \mu, e$. 
The latter case is not interesting for our analysis, 
which is related to constraints on $\tau \to V \ell$. 

\subsection{Unflavored pseudoscalar and scalar mesons.}
\label{sec:Pseudoscalar meson qq LFV decays}

The unflavored pseudoscalar and scalar mesons contribute to 
$\mu^{-} \to e^{-} \gamma \gamma$ and $\tau^{-}\to \ell^{-} \gamma\gamma$, 
according to the diagram in Fig.~\ref{fig:tau3ell}(b), 
with the LFV vertex $P(S) \bar\ell_{1}\ell_{2}$ 
given in Eq.~(\ref{eq:Meson-Lepton-LFV-Vert}), and 
\begin{eqnarray}
\label{eq:Vert-PGammaGamma}
{\cal L}_{P\gamma\gamma} &=& \frac{e^2}{4} \, g_{P\gamma\gamma} \, 
P \, 
F_{\mu\nu}\,  \varepsilon^{\mu\nu\alpha\beta}\,   F_{\alpha\beta}\,, \\ 
\label{eq:Vert-SGammaGamma}
{\cal L}_{S\gamma\gamma} &=& \frac{e^2}{4} \, g_{S\gamma\gamma} \, 
S\, F_{\mu\nu} \, F^{\mu\nu},
\end{eqnarray}
where
$F_{\mu\nu} = \partial_{\mu}A_{\nu}- \partial_{\nu}A_{\mu}$ 
is the electromagnetic stress tensor, $\varepsilon^{\mu\nu\alpha\beta}$
is the Levi-Cevita tensor, 
and $g_{I\gamma\gamma}$ ($I=P, S$) are 
the effective couplings of the $I \to \gamma\gamma$ 
decay widths: \eq 
\Gamma(I \to \gamma\gamma) = \frac{\pi\alpha^2}{4} \, 
g_{I\gamma\gamma}^2 \, M_I^3\,,  
\en  
where $\alpha \simeq 1/137.036$ is the fine structure constant. 
In the case of $\pi^0$, the coupling 
$g_{\pi\gamma\gamma}$ is related to the pion decay constant 
$F_\pi \simeq 92.4$ MeV as 
\eq 
g_{\pi\gamma\gamma} = \frac{1}{4 \pi^2 F_\pi}\,. 
\en 
The pion contribution to the decay $\mu \to e \gamma\gamma$ was 
discussed in Ref.~\cite{Nussinov:2000nm}. 
Extending this analysis to include other scalar and 
pseudoscalar mesons we can write
\begin{eqnarray}\label{eq:MU-EGG-1}
{\rm Br}(\mu^{-}\rightarrow e^{-} \gamma\gamma) &\approx&
\frac{\Gamma(I \to \mu  e) \,
\Gamma(I \to \gamma\gamma)}{\Gamma^2(W \to e \bar\nu_e)} \,
\left(\frac{M_W}{M_I}\right)^6 \left(\frac{M_{\mu}}{2 M_{I}}\right)^{4}.
\end{eqnarray}
For the $\tau$ lepton decay we find, in analogy  to Eq.~(\ref{eq:TauLEE-1}),
and for $M_{I}>M_{\tau}$:
\begin{eqnarray}\label{eq:Tau-EGG-1}
{\rm Br}(\tau^{-}\rightarrow \ell^{-} \gamma\gamma) &\approx&
\frac{\Gamma(I \to \tau \ell) \,
\Gamma(I \to \gamma\gamma)}{\Gamma^2(W \to e \bar\nu_e)} \,
\frac{\Gamma(\mu^{-}\to e^{-} \bar\nu_{e} \nu_{\mu})}{\Gamma(\tau \to All)}
\left(\frac{M_W}{M_I}\right)^6   \left(\frac{M_{\tau}}{M_{\mu}}\right)^5
  \left(\frac{M_{\tau}}{2 M_{I}}\right)^{4}\,.
\end{eqnarray}
The case $M_I < M_{\tau}$ with on-mass-shell mesons is not interesting
for our analysis.

In our numerical analysis, we use the central values of the decay widths 
of pseudoscalar and scalar mesons quoted from the Particle Data 
Group~\cite{PDG}: 
\eq 
& &
\Gamma(\pi^0   \to \gamma\gamma)  = 7.64   \ {\rm eV}\,, 
\quad \hspace{26pt}
\Gamma(\eta    \to \gamma\gamma)  = 0.52   \ {\rm keV}\,,  
\nonumber\\
& &
\Gamma(\eta'   \to \gamma\gamma)  = 4.35   \ {\rm keV}\,,  
\quad \hspace{24pt}
\Gamma(\eta_c  \to \gamma\gamma)  = 5.02   \ {\rm keV}\,,  
\nonumber\\ 
& &
\Gamma(f_0(500)\to \gamma\gamma)  = 2.05   \ {\rm keV}\,,  
\quad 
\Gamma(f_0(980)\to \gamma\gamma)  = 0.31   \ {\rm keV}\,,  
\nonumber\\
& &
\Gamma(a_0(980)\to \gamma\gamma)  = 0.30   \ {\rm keV}\,,  
\quad 
\Gamma(\chi_{c0}(1P)\to \gamma\gamma)  = 2.20 \ {\rm keV}\,. 
\en 
 
Note that up to now there are no experimental constraints on
$\tau \to \ell \gamma\gamma$ decay rates.
Therefore, in the present paper we present
only theoretical formula~(\ref{eq:Tau-EGG-1}) relating three-body LFV decay 
of $\tau$ with two-body LFV decays $P(S) \to \tau \ell$, which could be
useful in future searches of these processes. 

\subsection{Limits on two-body LFV meson decays}
\label{sec:Limits on two-body LFV meson decays}

From Eqs.~(\ref{eq:NussinovFivRes-1}), (\ref{eq:TauLEE-1}), 
(\ref{eq:MU-EGG-1}) and (\ref{eq:Tau-EGG-1}), 
we deduce upper limits for the branching ratios of the two-body LFV decays 
$M(Z) \to \ell_1 \ell_2$ of neutral vector and pseudoscalar mesons and 
$Z$-boson, using the existing data 
(\ref{eq:ExpLim-mu})-(\ref{eq:ExpLim-tau-mu}) for 
three-body LFV decays $\tau(\mu) \to 3 \ell$. 
We present our results in the second column of  
Table~\ref{tab:two-body-LFV} and compare them with the limits 
derived from the study of lepton conversion~\cite{Nussinov:2000nm}  
and available experimental data~\cite{PDG}. 

In the case of the $\pi^0$ and $J/\psi$ contributions, 
we also show in parenthesis our results for the constraints which
take into account the $Q^2$-dependence of the meson propagator 
and the form factor $\tilde{g}_{M\ell_{1}\ell_{2}}(Q^{2})$, when this last
effect is significant. For other meson contributions the effect of 
the $Q^{2}$-dependence is negligible. A detailed discussion and 
estimation of this effect is presented in Appendix~\ref{app:Q2dep}. 

\begin{table}[ht]
\begin{center}
\caption{Upper limits for the branching ratios of two-body LFV decays 
of neutral vector, pseudoscalar, and scalar mesons, and $Z$-boson, 
extracted from the bound on the indicated three-body $\mu$ and $\tau$ decays.
``EO-improved'' are limits obtained from Eq.~(\ref{eq:M1-M2-1}) 
relating different LFV processes with the same underlying 
effective operators (EO).} 
\label{tab:two-body-LFV}
\def\arraystretch{.9}
\begin{tabular}{|c|c|c|c|c|}
\hline 
Mode & \multicolumn{2}{|c|}{Our results} & Existing Limits & Data \\ 
\hline
\multicolumn{3}{|c|}{from $\mu^- \to e^- \gamma\gamma$ process}& &\\
\cline{1-3}
$\pi^0   \to \mu^\pm e^\mp$  
& \multicolumn{2}{|c|}{
$5.8 \times 10^{-11}$ ($3.2 \times 10^{-11}$)
}  
& $10^{-10}$~\cite{Nussinov:2000nm}
& $3.8 \times 10^{-10}$~\cite{PDG} \\
$\eta   \to \mu^\pm e^\mp$  
& \multicolumn{2}{|c|}{$6.2 \times 10^{-9}$} 
& $10^{-8}$~\cite{Nussinov:2000nm}
& $3.0 \times 10^{-6}$~\cite{PDG} \\
$\eta'   \to \mu^\pm e^\mp$  
& \multicolumn{2}{|c|}{$1.3 \times 10^{-9}$} 
& $4.7 \times 10^{-4}$~\cite{PDG} & \\
$\eta_c   \to \mu^\pm e^\mp$  
& \multicolumn{2}{|c|}{$5.9 \times 10^{-7}$}
& $1.57 \times 10^{-4}$~\cite{PDG} & \\
$f_0(500)   \to \mu^\pm e^\mp$  
& \multicolumn{2}{|c|}{$1.6 \times 10^{-15}$}
& &\\
$f_0(980)   \to \mu^\pm e^\mp$  
& \multicolumn{2}{|c|}{$1.0 \times 10^{-10}$}
& &\\
$a_0(980)   \to \mu^\pm e^\mp$  
& \multicolumn{2}{|c|}{$6.2 \times 10^{-11}$}
& &\\
$\chi_{c0}(1P)  \to \mu^\pm e^\mp$  
& \multicolumn{2}{|c|}{$1.5 \times 10^{-5}$}
& &\\
\hline\hline
Mode & Our results & EO-improved& Existing Limits & Data \\ 
\hline
\multicolumn{3}{|c|}{from $\mu^- \to e^- e^+ e^-$ process} &&\\
\cline{1-3}
$\rho^0   \to \mu^\pm e^\mp$  & $5.8 \times 10^{-21}$ & & 
$3.5 \times 10^{-24}$~\cite{Gutsche:2011bi}
& \\
$\omega   \to \mu^\pm e^\mp$  & $6.8 \times 10^{-20}$ 
& $9.1 \times 10^{-21}$ &
$6.2 \times 10^{-27}$~\cite{Gutsche:2011bi}
& \\ 
$\phi     \to \mu^\pm e^\mp$  & $1.6 \times 10^{-19}$ 
& $1.1 \times 10^{-19}$ &  
$4 \times 10^{-17}$~\cite{Nussinov:2000nm}; 
$(1.1 \times 10^{-25} - 5.6 
\times 10^{-22})$~\cite{Gutsche:2011bi}
& 
$2.0 \times 10^{-6}$~\cite{PDG} \\
$J/\psi   \to \mu^\pm e^\mp$  & $2.9 \times 10^{-17}$ 
& $2.6 \times 10^{-18}$ &  
$4 \times 10^{-13}$~\cite{Nussinov:2000nm}; 
$5.4 \times 10^{-14}$~\cite{Gutsche:2011bi}
& 
$1.6 \times 10^{-7}$~\cite{PDG} \\
$\Upsilon \to \mu^\pm e^\mp$  & $1.0 \times 10^{-13}$ 
&$2.5 \times 10^{-16}$ &  
$2 \times 10^{-9}$~\cite{Nussinov:2000nm}; 
$1.1 \times 10^{-6}$~\cite{Gutsche:2011bi}
& \\ 
$Z^0 \to \mu^\pm e^\mp$  & $1.3 \times 10^{-12}$ & & 
$5 \times 10^{-13}$~\cite{Nussinov:2000nm}; 
$4.2 \times 10^{-6}$~\cite{Gutsche:2011bi}
& 
$7.5 \times 10^{-7}$~\cite{PDG} \\
\hline
\multicolumn{3}{|c|}{ from $\tau^- \to e^- e^+ e^-$ process}&& \\
\cline{1-3}
$J/\psi   \to \tau^\pm e^\mp$ & 
$4.5 \times 10^{-12}$ ($2.8 \times 10^{-12}$) 
& & 
$6 \times 10^{-7}$~\cite{Nussinov:2000nm}  & 
$8.3 \times 10^{-6}$~\cite{PDG} \\ 
$\Upsilon \to \tau^\pm e^\mp$ & $1.6 \times 10^{-8}$  & $7.3 \times 10^{-10}$ &
$1 \times 10^{-2}$~\cite{Nussinov:2000nm}  & \\
$Z^0      \to \tau^\pm e^\mp$ & $1.9 \times 10^{-7}$  & & 
$3 \times 10^{-6}$~\cite{Nussinov:2000nm}  & 
$9.8 \times 10^{-6}$~\cite{PDG} \\
\hline
\multicolumn{3}{|c|}{from $\tau^- \to \mu^- e^+ e^-$ process}&& \\
\cline{1-3} 
$J/\psi   \to \tau^\pm \mu^\mp$ & 
$3.0 \times 10^{-12}$ ($1.9 \times 10^{-12}$) 
& &  & 
$2.0 \times 10^{-6}$~\cite{PDG} \\
$\Upsilon \to \tau^\pm \mu^\mp$ & $1.0 \times 10^{-8}$  
& $4.9 \times 10^{-10}$ & No limits& 
$6.0 \times 10^{-6}$~\cite{PDG} \\ 
$Z^0      \to \tau^\pm \mu^\mp$ & $1.3 \times 10^{-7}$  & &  & 
$1.2 \times 10^{-5}$~\cite{PDG} \\
\hline 
\end{tabular}
\end{center}
\end{table}

One can see from Table~\ref{tab:two-body-LFV} that in most cases 
we get more stringent constraints on 
the branching ratios of the two-body LFV decays. 
In particular, our limits are $3-4$ orders of magnitude better 
than the existing ones for $J/\Psi, \Upsilon \rightarrow \mu e$, 
while for $J/\Psi, \Upsilon \rightarrow \tau e$ the improvement is
5 orders of magnitude. 
To the best of our knowledge, in the 
literature there are no phenomenological limits  
for $J/\Psi, \Upsilon, Z \rightarrow \tau \mu$, and 
our limits are significantly more stringent than the existing 
experimental bounds~\cite{PDG}. 
In Table~\ref{tab:two-body-LFV},  
we also displayed for completeness the LFV decays of 
$f_{0},a_{0},\chi_{c0} \rightarrow \mu e$, which are unrealistic 
for experimental observations. We recall that these mesonic states, 
together with other mesons, are needed for the implementation of 
the quark-hadron duality and the derivation of the limits on 
the quark-lepton operators (\ref{eq:T-term-1}).


\section{Quark-lepton effective operators in LFV decays of $\mu, \tau$}
\label{sec:Lepto-quark effective LFV operators in LFV decays MU-TAU}

\subsection{Indirect contribution to $\ell_{1}\rightarrow \ell_{2} e e$}

Here we examine the limits on the effective quark-lepton
operators~(\ref{eq:l-q-operators}) from the purely leptonic 
processes \mbox{$\tau^{-}\rightarrow \mu^{-}(e^{-}) e^{+} e^{-}$}, 
$\mu^{-}\rightarrow e^{-} e^{+} e^{-}$ or 
$\tau^{-}\rightarrow \mu^{-}(e^{-}) \gamma\gamma$, 
$\mu^{-}\rightarrow e^{-} \gamma\gamma$.
The operators (\ref{eq:l-q-operators}) contribute to 
$\tau\rightarrow \ell e e$ 
at one-loop level. 
However, as we discussed 
in Sec.~\ref{sec:Effective Lepto-quark and Lepto-Meson LFV operators}, 
quark-hadron duality identifies these loop contributions with the tree-level 
contribution of the mesons states with the corresponding quantum numbers, 
as shown in  Fig.~\ref{fig:tau3ell}. 
In order to constrain the  quark-lepton operators~(\ref{eq:l-q-operators}), 
we match them to the corresponding meson-lepton operators 
in Eq.~(\ref{eq:Meson-Lepton-LFV-Vert}), using the on-mass-shell matching 
condition~\cite{Faessler:2004jt,Faessler:2005hx}: 
\begin{equation}\label{match1} 
\langle \ell_{1}^+ \, \ell_{2}^-|{\cal L}_{eff}^{lq}|M\rangle \approx 
\langle \ell_{1}^+ \, \ell_{2}^-|{\cal L}_{eff}^{lM}|M \rangle ,  
\end{equation} 
where $M$ are the corresponding  mass-shell meson states. 
This equation can be solved using the well-known quark current meson 
matrix elements shown in Appendix~\ref{App:A}, 
and we find relations between the quark-lepton scaled Wilson coefficients, 
$\mathcal{C}/\Lambda^{2}$ in Eq.~(\ref{eq:l-q-operators}), 
and the meson-lepton couplings, $g_{M}$, 
from~(\ref{eq:Meson-Lepton-LFV-Vert}), 
which are shown in Appendix~\ref{App:C}. 
Using these relations in the decay rate formulas 
for $\Gamma (M\rightarrow l_{1}l_{2})$ 
from Appendix A and substituting them into Eqs.~(\ref{eq:NussinovFivRes-1}),  
(\ref{eq:TauLEE-1}), (\ref{eq:MU-EGG-1}) and (\ref{eq:Tau-EGG-1}), 
we set upper limits on the coefficients $\mathcal{C}/\Lambda^{2}$ of 
the effective operators~(\ref{eq:l-q-operators}) from the experimental data on
$\tau(\mu) \to 3 \ell$. 
There are several operators contributing simultaneously to each of these 
processes, and therefore the data impose upper limits on linear combinations 
of the corresponding Wilson coefficients shown in Appendix~\ref{App:E}. 
In practice, it is useful to have individual upper limits for these 
coefficients under certain reasonable assumptions. In the literature,  
it is conventional to assume that there is no strong cancellation 
between terms of different origin in the amplitudes and therefore extract 
limits on each term as if it was present alone.  
We apply this ``one-at-a-time'' approach 
to Eqs.~(\ref{eq:gVVAV-Lambda2}-\ref{eq:gSSPS-Lambda2}). 
The corresponding results are displayed in 
Table~\ref{Table-2} in the form of  lower limits 
on the individual mass scales, $\Lambda^{ij}_{\mu e}$, 
of the operators in Eq.~(\ref{eq:l-q-operators}). 
In the conventional definition (see, for instance, 
Ref.~\cite{Gonzalez:2013rea}), these scales are
related to our notation as 
\begin{eqnarray}\label{Lambda-LFV}
|C^{XY}_{a}| \, \left(\frac{1{\rm GeV}}{\Lambda}\right)^2 = 
4\pi \left(\frac{1{\rm GeV}}{\Lambda^{XY}_{a}}\right)^2
\end{eqnarray}
with $a=0, 3, s, c,b, t$ and $z= hV, rS$, where $h = A, V$ and $r = P, S$ 
as defined before.   


\begin{table}

\vspace*{.4cm} 

\begin{center} 
\begin{tabular}{|c|c|c||c|c|c||c|c|c|} 
\hline
&&&&&&&&    \\[-3mm]
$\Lambda_{\ell_{1}\ell_{2}}$ & Our 
& Existing &$\Lambda_{\ell_{1}\ell_{2}}$ 
& Our & Existing & $\Lambda_{\ell_{1}\ell_{2}}$
& Our &Existing  \\
&limits &limits &&limits &limits&&limits  &limits\\
&[TeV] &[TeV] &&[TeV] &[TeV]&&[TeV]  &[TeV]\\[3mm]
\hline 
&& &&&&&&   \\[-3mm]
$\Lambda_{\mu e}^{(3) VV, AV}$ & 86 &   $10^{3}$  
&$\Lambda_{\mu e}^{(3) PP,SP} $ 
& 8.0&none&$\Lambda_{\tau e}^{(c) VV,AV} $&13&none  \\[1mm]
\hline
&& &&&&&&   \\[-3mm]
$\Lambda_{\mu e}^{(3) AA, VA} $ & 7.1 & none &$\Lambda_{\mu e}^{(s) SS, PS} $ 
& none& 
$3\times 10^{3}$ &$\Lambda_{\tau e}^{(b) VV, AV} $&7&none \\[1mm]
\hline
&&  &&&&&&  \\[-3mm] 
$\Lambda_{\mu e}^{(0) VV, AV} $ & 89 & $4.7 \times 10^{3}$  
&$\Lambda_{\mu e}^{(s) PP, SP} $ & 1.3& none   
&$\Lambda_{\mu e}^{(c) TT} $&19&none  \\[1mm]
\hline
&&   &&&&&& \\[-3mm]
$\Lambda_{\mu e}^{(0) AA, VA} $ & 2.4 & none 
&$\Lambda_{\mu e}^{(c) SS, PS} $ & none 
&   950 & $\Lambda_{\mu e}^{(b) TT} $&8.4&none \\[1mm]
\hline
&&  &&&&&&  \\[-3mm] 
$\Lambda_{\mu e}^{(s) VV,AV} $ & 134  &  $770$ 
& $\Lambda_{\mu e}^{(b) SS, PS} $ 
& none&  540 & $\Lambda_{\tau\mu}^{(c) VV, AV} $ &14.5&none  \\[1mm]
\hline
&&  &&&&&&  \\[-3mm] 
$\Lambda_{\mu e}^{(s) AA, VA} $ & 0.6 & none & $\Lambda_{\mu e}^{(t) SS, PS} $ 
&none&  90 & $\Lambda_{\tau\mu}^{(b) VV, AV} $&7.7&none \\[1mm]
\hline
&&  &&&&&&  \\[-3mm] 
$\Lambda_{\mu e}^{(c) VV, AV} $ & 300& 54   & $\Lambda_{\mu e}^{(3) TT} $ 
&103&none&$\Lambda_{\tau\mu}^{(c) TT} $ &19&none\\[1mm]
\hline
&&  &&&&&&  \\[-3mm] 
$\Lambda_{\mu e}^{(b) VV, AV} $& 138& 3   & $\Lambda_{\mu e}^{(0) TT} $ 
&107&none& $\Lambda_{\tau\mu}^{(b) TT} $&9.1&none\\[1mm]
\hline
&&  &&&&&&  \\[-3mm] 
$\Lambda_{\mu e}^{(3) SS, PS} $ & 0.5 &
$1.8 \times 10^{3}$  &$\Lambda_{\mu e}^{(s) TT} $&160&none&&&  \\[1mm]
\hline
&&  &&&&&&  \\[-3mm] 
$\Lambda_{\mu e}^{(0) SS, PS} $ & 0.6& $6.8 \times 10^{3} $  
& $\Lambda_{\mu e}^{(c) TT} $&355&none&&& \\[1mm]
\hline
&&  &&&&&&  \\[-3mm]
$\Lambda_{\mu e}^{(0) PP, SP} $ & 6.0& none   &$\Lambda_{\mu e}^{(b) TT} $ 
&164&none&&&\\[1mm]
\hline
\end{tabular}
\end{center} 
\caption{
Lower limits on the individual mass scales, $\Lambda_{\ell_{1}\ell_{2}}$, 
of the effective operators (\ref{eq:l-q-operators}).
``Existing limits'' are taken from Ref.~\cite{Faessler:2004jt}. 
All the limits  are derived assuming that only one operator contributes to 
$\tau\rightarrow \mu(e) e e$,  $\mu\rightarrow 3e$
at a time. 
}
\label{Table-2}
\end{table} 

\subsection{Relations between LFV decays of different mesons}
\label{sec:Relations between LFV decays of different mesons}

Notice that the operators in Eq.~(\ref{eq:l-q-operators}), either individually 
or in certain linear combination of them, underly LFV leptonic decay modes 
of all the mesons with the same quark content and  $J^{PC}$. 

Using the decay rate formulae, the meson matrix elements and the expressions 
for the LFV meson couplings from Appendices~\ref{App:A}, \ref{App:B} 
and \ref{App:C}, we find, in the limit of massless 
final leptons, the following approximate relation 
between the branching ratios of different mesons $\mathcal{M} =V,P$:
\begin{eqnarray}
\label{eq:M1-M2-1}
&&{\rm Br}(\mathcal{M}_{a}\rightarrow \ell_{1} \ell_{2}) \approx
\left(\frac{f_{{a}}}{f_{{b}}}\right)^{2}
\left(\frac{M_{{a}}}{M_{{b}}}\right)^{5}  
\frac{\Gamma(\mathcal{M}_{b}\rightarrow All)}{\Gamma(\mathcal{M}_{a}
\rightarrow All)}\cdot 
{\rm Br}(\mathcal{M}_{b}\rightarrow \ell_{1} \ell_{2}).
\end{eqnarray} 
Using this relation and the upper limits in Table~\ref{tab:two-body-LFV} 
on the branching ratios for one particular meson, we can set limits for 
the other ones.  
These ``cross-limits'', shown in the column ``EO improved''
of Table~\ref{tab:two-body-LFV},
are in some cases significantly more stringent than the limits 
derived directly from the contribution of 
the corresponding meson to $\tau\rightarrow \ell e e$. 


\section{Summary} 

We derived unitarity-inspired bounds on the two-body LFV 
decays of unflavored neutral vector and pseudoscalar mesons as well 
as of the $Z$-boson, from the experimental bounds 
on the leptonic LFV decays 
$\tau(\mu) \to  \ell e^{+}e^{-}, \ell \gamma\gamma$.  
Many of our limits are better than those existing to date in the literature. 
We also derived still nonexistent in the literature theoretical limits for 
$J/\Psi, \Upsilon, Z \rightarrow \tau \mu$, which are significantly more 
stringent that the experimental bounds.  
Using the fact that the LFV decays of 
the mesons with the same quark content and  $J^{PC}$ 
originate from the same linear combination of  quark-lepton operators,  
Eqs.~(\ref{eq:l-q-operators}), we derived improved limits on the decay rate 
of one meson from the more stringent limit of the decay rate of another meson. 
In some cases, this improvement approaches 3 orders of magnitude. 

We analyzed the contribution of 
quark-lepton operators~(\ref{eq:l-q-operators}) to purely leptonic processes 
$\tau(\mu) \to  \ell e^{+}e^{-}, \ell \gamma\gamma$, on the basis of 
the quark-hadron duality, which takes into account these contributions 
as coming from intermediate meson states. 
In this approach, the nonperturbative QCD effects in the quark loops 
are effectively considered by the meson masses and their 
leptonic decay constants  In order to realize this approach,  
we matched at the hadronization scale the quark-lepton and meson-lepton 
effective Lagrangians and derived relations between the quark- and 
meson-level effective LFV couplings. With this at hand, we extracted lower 
limits on the individual scales of many LFV operators 
from~(\ref{eq:l-q-operators}), 
which are shown in Table~\ref{Table-2}. The limits for the scales of the 
tensor, axial-vector, pseudoscalar operators, as well as 
for $(\bar{q} \Gamma q)(\bar{e} \Gamma \tau)$, 
$(\bar{q} \Gamma q)(\bar{\mu} \Gamma \tau)$  
are new, nonexisting in the literature.
These limits can be useful for LFV phenomenology, allowing model independent 
predictions for the LFV processes induced by the generic set of 
quark-lepton operators (\ref{eq:l-q-operators}).
 
\begin{acknowledgments}

This work was supported 
by the Carl Zeiss Foundation under Project ``Kepler Center f\"ur Astro- und
Teilchenphysik: Hochsensitive Nachweistechnik zur Erforschung des
unsichtbaren Universums (Gz: 0653-2.8/581/2),'' 
by Fondecyt (Chile) Grants No. 1150792, No. 1170171, No. 1180232 
and by CONICYT (Chile) Ring ACT1406, PIA/Basal FB0821, 
by the Russian Federation program ``Nauka'' (Contract No. 0.1764.GZB.2017), 
by the Tomsk State University Competitiveness Improvement Program under 
Grant No. 8.1.07.2018, and by the Tomsk Polytechnic University Competitiveness 
Enhancement Program (Grant No. VIU-FTI-72/2017).

\end{acknowledgments}

\appendix 

\section{Meson matrix elements}
\label{App:A}

Here we show the meson matrix elements needed for the matching between 
the quark and hadron levels of the effective theory used in our analysis. 
In the case of vector and scalar operators, these are 
\begin{eqnarray}\label{mat-el2} 
&&\langle 0 |\bar u \, \gamma_\mu \, u|\rho^0(p,\epsilon)\rangle \, = \, 
\, - \, \langle 0|\bar d \, \gamma_\mu \, d|\rho^0(p,\epsilon)\rangle 
= \, M_{\rho}^2 \, f_{\rho} \, \epsilon_\mu(p)\,,\\ 
&&\langle 0|\bar{u}\ \gamma_{\mu}\ u|\omega(p,\epsilon)\rangle\hspace{0.7mm} = 
\hspace{5mm} \langle 0|\bar{d}\ \gamma_{\mu}\ d|\omega(p,\epsilon)\rangle 
\hspace{4.8mm} = \, 3 \, M_{\omega}^2 \, f_{\omega} \, \epsilon_\mu(p)\,,\\ 
&&\langle 0|\bar{s}\ \gamma_{\mu}\ s|\phi(p,\epsilon)\rangle 
\hspace{2mm} = \, - \, 3 \, M_\phi^2 \, f_{\phi} \, \epsilon_\mu(p) \,,\\ 
&&\langle 0|\bar{c}\ \gamma_{\mu}\ c|J/\psi(p,\epsilon)\rangle 
\hspace{2mm} = \,  M_{J/\Psi}^2 \, f_{J/\psi} \, \epsilon_\mu(p) \,, \\ 
&&\langle 0|\bar{b}\ \gamma_{\mu}\ b|\Upsilon(p,\epsilon)\rangle 
\hspace{2mm} = \,  M_{\Upsilon}^2 \, f_{\Upsilon} \, \epsilon_\mu(p) \,,\\
&&\langle 0|\bar u \, u|f_0(p)\rangle \, = \, 
 \langle 0|\bar d \, d|f_0(p)\rangle 
\, = \, M_{f_0}^2 \, f_{f_0}\,, \label{f0-dec}
\\
&&     \langle 0|\bar u \, u|a_0(p)\rangle = 
\,- \, \langle 0|\bar d \, d|a_0(p)\rangle 
\, = \, M_{a_0}^2 \, f_{a_0} \,. 
\label{a0-dec} 
\end{eqnarray} 
Here $p$, $m_M$  and $f_{M}$ are the 4-momentum, mass and dimensionless 
decay constant of the meson $M$, respectively, and $\epsilon_{\mu}$ is 
the vector meson polarization state vector. 

The current central values of the meson decay constants $f_{V}$ 
and masses $m_V$ are~\cite{PDG}:
\begin{eqnarray}\label{constants-1}  
&&f_{\rho}   = 0.2,    \ \ 
  f_{\omega} = 0.059,  \ \ 
  f_{\phi}   = 0.074,  \ \    
  f_{J/\psi} = 0.134,  \ \  
  f_{\Upsilon} = 0.08, \ \ 
  f_{f_0} =  0.28\,, 
  f_{a_0} =  0.19\,, 
\\   
\label{constants-2} 
&&M_{\rho}   = 771.1  \,\, \mbox{MeV}, \,\,\,
  M_{\omega} = 782.6  \,\, \mbox{MeV}, \,\,\,  
  M_{\phi}   = 1019.5 \,\, \mbox{MeV}\,,\\
\label{constants-3} 
&& M_{J/\psi} = 3097\,  \mbox{MeV}, \ \ M_{\Upsilon} = 9460\, \mbox{MeV}\,, 
\ \  
M_{f_0} = 500 \,\,  \mbox{MeV},\ \  M_{a_0} = 980 \,\, \mbox{MeV} 
\,.   
\end{eqnarray} 
The decay constants $f_{f_{0}}$ and $f_{a_{0}}$  
in Eqs.~(\ref{f0-dec}) and~(\ref{a0-dec}) 
are not yet known experimentally. The value $f_{f_{0}}$ 
was evaluated in Ref.~\cite{Faessler:2005hx} in the linear $\sigma$-model,,  
using the approach of Refs.~\cite{Delbourgo:1993dk,Delbourgo:1998ji} and 
the value $a_{f_{0}}$ was estimated using QCD sum rules~\cite{Maltman:1999jn}.

In the evaluation of tensor operators, we use the identity 
\eq\label{eq:sigma-0}
\sigma^{\mu\nu} \gamma^5 = \frac{i}{2} \, \varepsilon^{\mu\nu\alpha\beta} 
\sigma_{\alpha\beta}\,, 
\en 
which simplifies/constrains the structure of effective 
Lagrangians with tensor spin structure as 
\eq 
\label{eq:sigma-1}
& &\bar \ell_1 \sigma^{\mu\nu} P_{L/R} \ell_2 \, 
   \bar q_f    \sigma_{\mu\nu} P_{L/R} q_i    
 =\frac{1}{2}
   \bar \ell_1 \sigma^{\mu\nu} \ell_2 
   \bar q_f    \sigma_{\mu\nu} q_i    \,, \\
\label{eq:sigma-2}   
& &\bar \ell_1 \sigma^{\mu\nu} P_{L/R} \ell_2 \, 
   \bar q_f    \sigma_{\mu\nu} P_{R/L} q_i \equiv 0\,. 
\en 

The matrix element of the tensor quark operator is calculated according to 
\eq 
\langle 0| \bar q_f \, \sigma_{\mu\nu} \, q_i | V(p,\epsilon) \rangle \, = \, 
i \, (m_i + m_f) \, 
\Big(\epsilon_\mu(p) p_\nu - \epsilon_\nu(p) p_\mu\Big) \, f_V \,.
\en 
In deriving effective Lagrangians with derivates acting on meson fields,  
we use the convention that the meson is described by an incoming plane wave 
of the form $e^{-ipx}$. Therefore, the correspondence between the Lorentz 
structure $\epsilon_\mu(p) p_\nu - \epsilon_\nu(p) p_\mu$ 
and the field tensor of a vector meson in coordinate space is set as  
$i (\epsilon_\mu(p) p_\nu - \epsilon_\nu(p) p_\mu) \to F_{\mu\nu}(x)$. 

In the calculation of matrix elements of 
pseudoscalar, axial, and pseudotensor quark operators, 
we use the well-known  
relations~\cite{Gasser:1983yg,Gasser:1984gg,Gasser:1982ap} 
\eq 
& &\langle 0|\bar q_f \gamma^\mu\gamma^5 q_i |P(p) \rangle =
i p^\mu F_P  \,, \\
& &\langle 0|\bar q_f i \gamma^5 q_i |P(p) \rangle =
\frac{M_P^2}{m_i + m_f} \, F_P\,, 
\en 
where the $P$ meson has flavor structure $P=(q_i\bar q_f)$, 
$f_P$ is the pseudoscalar meson coupling constants. 
In the case of pseudoscalar mesons, we introduce singlet-octet mixing,
with a mixing angle of $\theta_{P} = -13.34^\circ$~\cite{Ambrosino:2006gk} 
\eq
\eta &\longrightarrow&
-\,\tfrac{1}{\sqrt{2}}\,\sin\delta\, (\bar u u + \bar d d)
-\,\cos\delta\, \bar s s \,,
\nonumber\\
\eta' &\longrightarrow&
+\,\tfrac{1}{\sqrt{2}}\,\cos\delta\, (\bar u u + \bar d d)
-\,\sin\delta\, \bar s s \,,
\nonumber\\ 
\delta &=& \theta_P-\theta_I, \qquad \theta_I=\arctan{\tfrac{1}{\sqrt{2}}}\,.
\label{eq:mixing}
\en 
The masses of the pseudoscalar mesons used in our calculations 
are~\cite{PDG}
\eq 
\label{eq:PS-masses}
& &
M_{\pi^0} = 134.977 \pm 0.0005\, \mbox{MeV}\,, \ \ \ 
M_{\eta}  = 547.862 \pm 0.017 \, \mbox{MeV}\,, \ \ \ 
M_{\eta'} = 957.78  \pm 0.06  \, \mbox{MeV}\,, \nonumber\\
& &
M_{\eta_c}= 2983.9  \pm 0.5   \, \mbox{MeV}\,. 
\en
For the pseudoscalar decay constants of $\pi^0$, $\eta$,  
and $\eta'$ mesons we use the universal value 
identified with the pion coupling 
$F_\pi = 92.4$ MeV. For the $\eta_c$ coupling we 
take the averaged value of theoretical predictions 
$F_{\eta_c} = 285$ MeV from Ref.~\cite{Gutsche:2018utw}. 

Therefore, the matrix elements of specific pseudoscalar 
and axial operators between vacuum and pseudoscalar states 
are: 
\eq 
& &\langle 0|\bar u \gamma^\mu\gamma^5 u |\pi^0(p) \rangle =
 - \langle 0|\bar d\gamma^\mu\gamma^5 d |\pi^0(p) \rangle =
i p^\mu F_\pi  \,, \\
& &\langle 0|\bar u \gamma^\mu\gamma^5 u |\eta(p) \rangle =
   \langle 0|\bar d \gamma^\mu\gamma^5 d |\eta(p) \rangle =
- i p^\mu F_\pi \, \sin\delta \,, \\
& &\langle 0|\bar u \gamma^\mu\gamma^5 u |\eta'(p) \rangle =
   \langle 0|\bar d \gamma^\mu\gamma^5 d |\eta'(p) \rangle =
  i p^\mu F_\pi \, \cos\delta \,, \\
& &\langle 0|\bar s \gamma^\mu\gamma^5 s |\eta(p) \rangle =
- i p^\mu F_\pi \, \cos\delta \sqrt{2} \,, \\
& &\langle 0|\bar s \gamma^\mu\gamma^5 s |\eta'(p) \rangle =
- i p^\mu F_\pi \, \sin\delta \sqrt{2} \,, \\
& &\langle 0|\bar c \gamma^\mu\gamma^5 c |\eta_c(p) \rangle =
i p^\mu F_{\eta_c}  \,, \\
& &\langle 0|\bar u i \gamma^5 u |\pi^0(p) \rangle =
 - \langle 0|\bar d i \gamma^5 d |\pi^0(p) \rangle =
\frac{M_\pi^2}{2 \hat{m}} \, F_\pi\,, \\ 
& &\langle 0|\bar u i \gamma^5 u |\eta(p) \rangle =
   \langle 0|\bar d i \gamma^5 d |\eta(p) \rangle =
- \frac{M_\eta^2}{2 \hat{m}} \, F_\pi \, \sin\delta \,, \\
& &\langle 0|\bar u i \gamma^5 u |\eta'(p) \rangle =
   \langle 0|\bar d i \gamma^5 d |\eta'(p) \rangle =
   \frac{M_{\eta'}^2}{2 \hat{m}} \, F_\pi \, \cos\delta \,, \\
& &\langle 0|\bar s i \gamma^5 s |\eta(p) \rangle =
-    \frac{M_{\eta}^2}{2 \hat{m}} \, F_\pi \, \cos\delta \sqrt{2}\,, \\
& &\langle 0|\bar s i \gamma^5 s |\eta'(p) \rangle =
-    \frac{M_{\eta'}^2}{2 \hat{m}} \, F_\pi \, \sin\delta \sqrt{2}\,, \\
& &\langle 0|\bar c i \gamma^5 c |\eta_c(p) \rangle = 
\frac{M_{\eta_c}^2}{2 m_c} \, F_{\eta_c}\,, 
\en 
where $\hat{m} = m_u = m_d = 7$ MeV is the mass of $u$ and $d$ 
quarks in the isospin limit, $m_s = 25 \hat{m}$ is the strange quark 
mass~\cite{Gasser:1982ap}, $m_c = 1.275$ GeV 
and $m_b = 4.18$ GeV are the masses of charm and bottom 
quarks~\cite{PDG}.

\section{LFV rates  of mesons decaying into leptonic pair.} 
\label{App:B}

Here, we present analytical results for the LFV rates of mesons 
decaying into a leptonic pair governed by the effective 
Lagrangian~(\ref{eq:Meson-Lepton-LFV-Vert}) and including 
effects of finite lepton masses, 

$V \to \ell_1^+ \ell_2^-$ decays 
\eq
\label{eq:rate-V-1} 
\Gamma(V \to \ell_1^+ \ell_2^-) &=& 
\frac{P^\ast}{6 \pi} \, 
  \biggl[
\biggl(g_{V \ell_1\ell_2}^{(V)}\biggr)^2 \ 
\biggl( 1 - \frac{M_-^2}{M_V^2} \biggr) \, 
\biggl( 1 + \frac{M_+^2}{2M_V^2} \biggr)  \, + \, 
\biggl(g_{V \ell_1\ell_2}^{(A)}\biggr)^2 \ 
\biggl( 1 - \frac{M_+^2}{M_V^2} \biggr) \, 
\biggl( 1 + \frac{M_-^2}{2M_V^2} \biggr)  \nonumber\\
&+& 
2 \biggl(g_{V \ell_1\ell_2}^{(T)}\biggr)^2 \ 
\biggl( 1 - \frac{M_-^2}{M_V^2} \biggr) \, 
\biggl( 1 + \frac{2M_+^2}{M_V^2} \biggr)  \, - \, 
6 g_{V \ell_1\ell_2}^{(V)} \, g_{V \ell_1\ell_2}^{(A)} \, 
\frac{M_+}{M_V} \, \biggl( 1 - \frac{M_-^2}{M_V^2} \biggr) 
\biggr]
\en 
$S \to \ell_1^+ \ell_2^-$ decays 
\eq
\Gamma(S \to \ell_1^+ \ell_2^-) = 
\frac{P^\ast}{4 \pi} \, 
\biggl[\biggl(g_{S\ell_1\ell_2}^{(S)}\biggr)^2 
\, \biggl(1 - \frac{M_-^2}{M_P^2}\biggr)
\,+\, 
\biggl(g_{S\ell_1\ell_2}^{(P)}\biggr)^2 
\, \biggl(1 - \frac{M_+^2}{M_P^2}\biggr)
\biggr] 
\en 
$P \to \ell_1^+ \ell_2^-$ decays 
\eq
\Gamma(P \to \ell_1^+ \ell_2^-) = 
\frac{P^\ast}{4 \pi} \, 
\biggl[\biggl(g_{P\ell_1\ell_2}^{(P)} + g_{P\ell_1\ell_2}^{(A)} \, 
\frac{M_+}{M_P}\biggr)^2 \, \biggl(1 - \frac{M_-^2}{M_P^2}\biggr) 
\,+\, 
\biggl(g_{P\ell_1\ell_2}^{(S)} + g_{P\ell_1\ell_2}^{(V)} \, 
\frac{M_-}{M_P}\biggr)^2 \, \biggl(1 - \frac{M_+^2}{M_P^2}\biggr) 
\biggr] 
\en 
where $M_\pm = M_{\ell_1} \pm M_{\ell_2}$, 
$P^\ast = \lambda^{1/2}(M_H^2,M_{\ell_1}^2,M_{\ell_2}^2)/(2 M_H)$ is 
the magnitude of the three momentum of leptons in the rest 
frame of decaying hadron $H$ and 
$\lambda(x,y,z) = x^2 + y^2 + z^2 - 2 xy - 2 yz - 2 xz$ 
is the kinematical triangle K\"allen function. 

\section{$Q^2$ dependence of meson propagators and form factors}
\label{app:Q2dep}

Let us note that in Eqs.~(\ref{eq:mu-3e-1}), (\ref{eq:mu-3e-2}),
(\ref{eq:TauLEE-1}), (\ref{eq:MU-EGG-1}), and (\ref{eq:Tau-EGG-1}), 
we neglected the squared momentum transfer
$Q^{2}$-dependence of the meson propagator and the form factors
$\tilde{g}_{M \ell_{1}\ell_{2}}(Q^{2})$.
For most of the processes of our current interest, this $Q^{2}$-dependence
results in a less than $5\%$ deviation from the approximate formulae 
that we use, which for our purposes is more than sufficient. 
Nevertheless, for two specific states (the intermediate pion in the process
$\mu \to e \gamma\gamma$ and the intermediate $J/\psi$ in the processes 
$\tau\to e(\mu) e e$) the $Q^2$ dependence of the meson propagator and 
of the form factors give contributions up to 80\%.

Here, we present details of the $Q^2$ dependent contribution calculation 
of the meson form factors $\tilde{g}_{M \ell_{1}\ell_{2}}(Q^{2})$
and propagators $D_M(Q^2)$ to the branchings of the three-body
LFV decays of leptons. The meson form factors
$\tilde{g}_{M \ell_{1}\ell_{2}}(Q^{2})$ can be found using a
covariant confined quark model~\cite{Branz:2009cd}. Their $Q^2$
dependence can be parametrized as
\eq
\tilde{g}_{M \ell_{1}\ell_{2}}(Q^{2}) = 1/(1-Q^2/\Lambda_M^2)\,,
\en
where $\Lambda_M$ is the set of cutoff parameters given by 
\eq
& &
\Lambda_\pi     = 0.90 \ {\rm GeV}\,, \quad
\Lambda_\eta    = 0.94 \ {\rm GeV}\,, \quad
\Lambda_{\eta'} = 1.02 \ {\rm GeV}\,, \quad
\Lambda_{\eta_c}= 4.16 \ {\rm GeV}\,, \nonumber\\
& &
\Lambda_{f_0(500)} = 1.02 \ {\rm GeV}\,, \quad
\Lambda_{f_0(980)} = 1.04 \ {\rm GeV}\,, \quad
\Lambda_{a_0(980)} = 1.06 \ {\rm GeV}\,, \quad
\Lambda_{\chi_{c0}}= 5.95 \ {\rm GeV}\,, \\
& &
\Lambda_{\rho}    =  0.84 \ {\rm GeV}\,, \quad
\Lambda_{\omega}  =  0.83 \ {\rm GeV}\,, \quad
\Lambda_{\phi}    =  1.13 \ {\rm GeV}\,, \quad
\Lambda_{J/\psi}  =  3.54 \ {\rm GeV}\,, \quad
\Lambda_{\Upsilon}   = 10.07 \ {\rm GeV}\,.
\nonumber
\en
In particular, we parametrize this effect by a factor $R$, which is defined
as the ratio of the three LFV decay branching with 
the complete $Q^2$ dependence (full result) and the branching
without that dependence: 
\eq
R = \frac{{\rm Br}_{\rm full}(\ell_1 \rightarrow \ell_2 + X)}
{{\rm Br}(\ell_1 \rightarrow \ell_2 + X)} \,.
\en
The coefficient $R$ is simply the ratio of the phase space
integrals for three-body LFV decays of leptons including
the form factors $I_{\rm full}$ and without such effects $I$ 
\eq
R = \frac{I^{\rm phase}_{\rm full}}{I^{\rm phase}}\,,
\en
where
\eq
I^{\rm phase}_{\rm full} &=& \frac{\pi^2}{4 M_1^2} \,
\int\limits_{s_2^-}^{s_2^+} ds_2 \, \lambda^{1/2}(M_1^2,s_2,M_2^2) \,
\tilde{g}_{M \ell_{1}\ell_{2}}^2(s_2) \, D_M^2(s_2) \,, \\
I^{\rm phase} &=& \frac{\pi^2}{4 M_1^2} \, \frac{1}{M^4} \,
\int\limits_{s_2^-}^{s_2^+} ds_2 \, \lambda^{1/2}(M_1^2,s_2,M_2^2) \,.
\en
Here $D_M(s_2) = 1/(M^2 - s_2)$ is the scalar part of meson propagator,
$s_2$ is the Mandelstam variable (invariant mass of two-lepton or
two-photon pair in the final state). The upper $(s_2^+)$ and lower
$(s_2^-)$ limits of the $s_2$ variation  are defined in terms of 
the initial lepton masses ($M_1$), final lepton masses $(M_2)$ and masses of
the leptonic pair ($M_3$,$M_4$) produced by the intermediate meson, as 
$s_2^+ = (M_1-M_2)^2$ and $s_2^- = (M_3+M_4)^2$.  
In the case of two-photon processes $s_2^- = 0$.
In the evaluation of $R_{\rm pr}$ and $R_{\rm ff}$, we
drop the $Q^2$ dependence of the meson propagator
$D_M(s_2) \to 1/M^2$ or the meson form factor
$\tilde{g}_{M \ell_{1}\ell_{2}}^2(s_2) \to 1$, respectively.

In Table~\ref{tab:Q2dep}, we explicitly
demonstrate the effect on the three-body LFV decay rates
of the $Q^2$-dependence of the meson propagator and form factors.
In particular, we parametrize this effect by the factor $R$, which is defined
as the ratio of the three-body LFV decay taking into account
the $Q^2$ dependences (full result) and the decay
without that dependence. 
We present separate results coming from the $Q^2$ dependence in the meson
propagators (factor $R_{\rm pr}$) and in the form factors (factor $R_{\rm ff}$)
and also the total results (factor $R$) combining these two contributions. 
From Table~\ref{tab:Q2dep}
one can see that effects of form factors are suppressed
for all processes and mesons and less 2\% except $\tau$ decays with
$J/\psi$ meson in the intermediate state giving about 20\% contribution.
$Q^2$ dependence of meson propagators is less than 3\% for most cases
except $\sim 80\%$ contribution of $\pi^0$ to the
${\rm Br}(\mu^- \to e^- \gamma \gamma)$ and
$\sim 30\%$ contribution of $J/\psi$ to the
${\rm Br}(\tau^- \to \ell^- e^+ e^-)$.
It is clear that the sizeable factors $R$ due to the $Q^2$ dependence
in case of mentioned mesons and modes give more stringent constraints
on two-body LFV meson decays.

\begin{table}[hb]
\begin{center}
\caption{Factors $R$, $R_{\rm pr}$, and $R_{\rm ff}$ representing
$Q^2$ dependence.}
\label{tab:Q2dep}
\def\arraystretch{.9}
\begin{tabular}{|c|c|c|c|}
\hline
Meson & \multicolumn{3}{|c|}{$\mu^- \to e^- \gamma\gamma$ process} \\
\cline{2-4}
      & \ \ $R_{\rm pr}$ \ \ & \ \ $R_{\rm ff}$ \ \ & $R$ \\
\hline
$\pi^0 $  & 1.788  & 1.009 & 1.808  \\
$\eta  $  & 1.025  & 1.008 & 1.034  \\
$\eta' $  & 1.008  & 1.007 & 1.015  \\
$\eta_c$  & 1.0008 & 1.0004& 1.0013 \\
$f_0(500)$& 1.031  & 1.007 & 1.038  \\
$f_0(980)$& 1.008  & 1.007 & 1.015  \\
$a_0(980)$& 1.008  & 1.007 & 1.015  \\
$\chi_{c0}(1P)$& 1.0006 & 1.0002 & 1.0013 \\
\hline
Meson    & \multicolumn{3}{|c|}{$\mu^- \to e^- e^+ e^-$ process} \\
\cline{2-4}
      & \ \ $R_{\rm pr}$ \ \ & \ \ $R_{\rm ff}$ \ \ & $R$ \\
\hline
$\rho^0 $  & 1.013 & 1.011 & 1.023 \\
$\omega $  & 1.012 & 1.011 & 1.023 \\
$\phi   $  & 1.007 & 1.006 & 1.013 \\
$J/\psi $  & 1.0008 & 1.0006 & 1.001 \\
$\Upsilon$ & 1.0001 & 1.0001 & 1.0002\\
\hline
Meson      & \multicolumn{3}{|c|}{$\tau^- \to e^- e^+ e^-$ process} \\
\cline{2-4}
      & \ \ $R_{\rm pr}$ \ \ & \ \ $R_{\rm ff}$ \ \ & $R$ \\
\hline
$J/\psi$  & 1.293 & 1.208 & 1.605 \\
$\Upsilon$& 1.024 & 1.021 & 1.045 \\
\hline
Meson      & \multicolumn{3}{|c|}{$\tau^- \to \mu^- e^+ e^-$ process} \\
\cline{2-4}
      & \ \ $R_{\rm pr}$ \ \ & \ \ $R_{\rm ff}$ \ \ & $R$ \\
\hline
$J/\psi$  & 1.273 & 1.195 & 1.555 \\
$\Upsilon$& 1.023 & 1.020 & 1.044 \\
\hline
\end{tabular}
\end{center}
\end{table}

\section{Relations of meson-lepton to quark-lepton couplings}
\label{App:C}

Here we show the relation between quark-lepton, $\mathcal{C}_{qq}$, 
and meson-lepton, $g_{M}$, couplings from Eqs.~(\ref{eq:l-q-operators}) 
and~(\ref{eq:Meson-Lepton-LFV-Vert}) 
derived as solutions of the matching conditions (\ref{match1}).  
They are as follows
\begin{eqnarray}\label{eq:gVA-Lambda}
g^{(V/A)}_{\rho^0 \ell_1\ell_2} &=& \frac{M_\rho^2}{\Lambda^2} 
\, f_{\rho}  \, \mathcal{C}^{(3)VV/AV}_{\ell_1\ell_2}
\,,\quad 
g^{(V/A)}_{\omega \ell_1\ell_2} \, = \,  \frac{3 M_\omega^2}{\Lambda^2} 
\, f_{\omega}  \, \mathcal{C}^{(0)VV/AV}_{\ell_1\ell_2}
\,,\quad 
g^{(V/A)}_{\phi \ell_1\ell_2} \, = \,  - \frac{3 M_\phi^2}{\Lambda^2} 
\, f_{\phi}  \, \mathcal{C}^{(s) VV/AV}_{\ell_1\ell_2}
\,,\nonumber\\ 
g^{(V/A)}_{J/\psi \ell_1\ell_2} &=& \frac{M_{J/\psi}^2}{\Lambda^2} 
\, f_{J/\psi}  \, \mathcal{C}^{(c) VV/AV}_{\ell_1\ell_2}
\,,\quad 
g^{(V/A)}_{\Upsilon \ell_1\ell_2} \, = \,  \frac{M_{\Upsilon}^2}{\Lambda^2} 
\, f_{\Upsilon}  \, \mathcal{C}^{(b) VV/AV}_{\ell_1\ell_2}
\,,\nonumber\\ 
g^{(T)}_{\rho \ell_1\ell_2} &=&  
\frac{\hat{m} M_{\rho}}{\Lambda^{2}} \, f_{\rho} \, 
\mathcal{C}^{(3) TT}_{\ell_1\ell_2} 
\,, \quad 
g^{(T)}_{\omega \ell_1\ell_2} \, = \, 
\frac{3 \hat{m} M_{\omega}}{\Lambda^{2}} \, f_{\omega} \, 
\mathcal{C}^{(0) TT}_{\ell_1\ell_2} 
\,, \quad 
g^{(T)}_{\phi\ell_1\ell_2} \, = \,  
- \frac{3 m_s M_{\phi}}{\Lambda^{2}} \, f_{\phi} 
\mathcal{C}^{(s) TT}_{\ell_1\ell_2} 
\,, \nonumber\\
g^{(T)}_{J/\psi\ell_1\ell_2} &=&
\frac{m_c M_{J/\psi}}{\Lambda^{2}} \, f_{J/\psi} 
\mathcal{C}^{(c) TT}_{\ell_1\ell_2} 
\,, \quad 
g^{(T)}_{\Upsilon\ell_1\ell_2} \, = \,   
\frac{m_b M_{\Upsilon}}{\Lambda^{2}} \, f_{\Upsilon} 
\mathcal{C}^{(b) TT}_{\ell_1\ell_2} 
\,, \label{Eq:relations}\\
g^{(S/P)}_{a_0 \ell_1\ell_2} &=& 
\frac{M_{a_0}^2}{\Lambda^{2}} \, f_{a_0} \,
\mathcal{C}^{(3) SS/PS}_{\ell_1\ell_2} 
\,,\quad 
g^{(S/P)}_{f_0 \ell_1\ell_2} \, = \,  
\frac{M_{f_0}^2}{\Lambda^{2}} \, f_{f_0} \,  
\mathcal{C}^{(0) SS/PS}_{\ell_1\ell_2} 
\,,\quad 
g^{(S/P)}_{\chi_{c0} \ell_1\ell_2} \, = \,
\frac{M_{\chi_{c0}}^2}{\Lambda^{2}} \, f_{f_0} \,
\mathcal{C}^{(c) SS/PS}_{\ell_1\ell_2}
\,, 
\nonumber\\
g^{(P/S)}_{\pi \ell_1\ell_2} &=&  
\frac{M_\pi^2}{2 \hat{m} \Lambda^2} \, F_\pi \, 
\mathcal{C}^{(3) PP/SP}_{\ell_1\ell_2} 
\,,\quad 
g^{(P/S)}_{\eta_c \ell_1\ell_2} \, = \,
\frac{M_{\eta_c}^2}{2 m_c \Lambda^2} \, F_{\eta_c} \, 
\mathcal{C}^{(c) PP/SP}_{\ell_1\ell_2} 
\,,\nonumber\\ 
g^{(P/S)}_{\eta \ell_1\ell_2} &=& 
- \frac{M_\eta^2}{2 \hat{m} \Lambda^2} \, F_\pi \, 
\biggl( \sin\delta \,
\mathcal{C}^{(0) PP/SP}_{\ell_1\ell_2} 
+ \frac{\hat{m}}{m_s} \, \cos\delta \, \sqrt{2} \, 
\mathcal{C}^{(s) PP/SP}_{\ell_1\ell_2} 
\biggr)
\,,
\nonumber\\
g^{(P/S)}_{\eta' \ell_1\ell_2} && 
\frac{M_{\eta'}^2}{2 \hat{m} \Lambda^2} \, F_\pi \,
\biggl( \cos\delta \, 
\mathcal{C}^{(0) PP/SP}_{qq^+, \ell_1\ell_2} 
- \frac{\hat{m}}{m_s} \, \sin\delta \, \sqrt{2} \, 
\mathcal{C}^{(s) PP/SP}_{\ell_1\ell_2} 
\biggr) 
\,,\nonumber\\ 
g^{(A/V)}_{\pi \ell_1\ell_2} &=&  
- \frac{M_\pi}{\Lambda^2} \, F_\pi \, 
\mathcal{C}^{(3) AA/VA}_{\ell_1\ell_2} 
\,,\quad 
g^{(A/V)}_{\eta_c \ell_1\ell_2} \, = \, 
- \frac{M_{\eta_c}}{\Lambda^2} \, F_{\eta_c} \, 
\mathcal{C}^{(c) AA/VA}_{\ell_1\ell_2} \,,
\nonumber\\
g^{(A/V)}_{\eta \ell_1\ell_2} &=&    
 \frac{M_\eta}{\Lambda^2} \, F_\pi \, 
\biggl( \sin\delta \,
\mathcal{C}^{(0) AA/VA}_{\ell_1\ell_2} 
\,+\, \cos\delta \, \sqrt{2} \, 
\mathcal{C}^{(s) AA/VA}_{\ell_1\ell_2} 
\biggr)
\,,
\nonumber\\
g^{(A/V)}_{\eta' \ell_1\ell_2} &=&  
- \frac{M_{\eta'}}{\Lambda^2} \, F_\pi \, 
\biggl( \cos\delta \,
\mathcal{C}^{(0) AA/VA}_{\ell_1\ell_2} 
\,-\, \sin\delta \, \sqrt{2} \, 
\mathcal{C}^{(0) AA/VA}_{\ell_1\ell_2} 
\biggr) \,, 
\nonumber
\en
where 
\eq 
\mathcal{C}^{(0/3) \Gamma_i\Gamma_J}_{\ell_1\ell_2} = 
\mathcal{C}^{(u) \Gamma_i\Gamma_J}_{\ell_1\ell_2} \pm 
\mathcal{C}^{(d) \Gamma_i\Gamma_J}_{\ell_1\ell_2} 
\en 
is the strong isospin singlet $\mathcal{C}^{(0)}$ and $\mathcal{C}^{(3)}$ 
triplet combinations.

\section{Limits on linear combinations of the Wilson coefficients}
\label{App:E}

Here, we show the limits on the combinations of the quark-lepton couplings
$\bar{\mathcal{C}} = \mathcal{C} \cdot (1\,{\rm TeV}/\Lambda)^2$ are
\eq
\label{eq:gVVAV-Lambda2}
& &
\left\vert
\mathcal{C}^{(3)VV/AV}_{\mu e} 
\right\vert < 1.7 \times 10^{-3}\,, \quad 
\left\vert
\mathcal{C}^{(0)VV/AV}_{\mu e} 
\right\vert < 1.6 \times 10^{-3}\,, \quad 
\left\vert
\mathcal{C}^{(s)VV/AV}_{\mu e} 
\right\vert < 0.7 \times 10^{-3}\,, \\
& &
\left\vert
\mathcal{C}^{(c)VV/AV}_{\mu e} 
\right\vert < 1.4 \times 10^{-4}\,, \quad 
\left\vert
\mathcal{C}^{(c)VV/AV}_{\tau e} 
\right\vert < 7.6 \times 10^{-2}\,, \quad 
\left\vert
\mathcal{C}^{(c)VV/AV}_{\tau \mu} 
\right\vert < 6.0 \times 10^{-2}\,, \\                                       
& &
\left\vert
\mathcal{C}^{(b)VV/AV}_{\mu e}   
\right\vert < 6.6 \times 10^{-4}\,, \quad
\left\vert
\mathcal{C}^{(b)VV/AV}_{\tau e} 
\right\vert < 2.6 \times 10^{-1}\,, \quad 
\left\vert
\mathcal{C}^{(b)VV/AV}_{\tau \mu} 
\right\vert < 2.1 \times 10^{-1}\,, 
\en
\eq
\label{eq:gTT-Lambda2}
& &
\left\vert
\mathcal{C}^{(3)TT}_{\mu e} 
\right\vert < 1.2 \times 10^{-3}\,, \quad 
\left\vert
\mathcal{C}^{(0)TT}_{\mu e} 
\right\vert < 1.1 \times 10^{-3}\,, \quad 
\left\vert
\mathcal{C}^{(s)TT}_{\mu e} 
\right\vert < 0.5 \times 10^{-3}\,, \\
& &
\left\vert
\mathcal{C}^{(c)TT}_{\mu e} 
\right\vert < 1.0 \times 10^{-4}\,, \quad 
\left\vert
\mathcal{C}^{(c)TT}_{\tau e} 
\right\vert < 4.5 \times 10^{-2}\,, \quad 
\left\vert
\mathcal{C}^{(c)TT}_{\tau \mu} 
\right\vert < 3.5 \times 10^{-2}\,, \\
& &
\left\vert
\mathcal{C}^{(b)TT}_{\mu e}  
\right\vert < 4.7 \times 10^{-4}\,, \quad 
\left\vert
\mathcal{C}^{(b)TT}_{\tau e} 
\right\vert < 1.8 \times 10^{-2}\,, \quad 
\left\vert
\mathcal{C}^{(b)TT}_{\tau \mu} 
\right\vert < 1.5 \times 10^{-2}\,, 
\en
\eq
\label{eq:gPPSP-Lambda2}
& &
\left\vert
\mathcal{C}^{(3)PP/SP}_{\mu e}    
\right\vert < 0.2 \,, \quad 
\left\vert
\mathcal{C}^{(c)PP/SP}_{\mu e}    
\right\vert < 4.0 \times 10^{2} \,, \\
& &
\left\vert
\mathcal{C}^{(0)PP/SP}_{\mu e} 
- 0.05 \, \mathcal{C}^{(s)PP/SP}_{\mu e} 
\right\vert < 0.4\,, \quad 
\left\vert
\mathcal{C}^{(0)PP/SP}_{\mu e} 
+ 0.06 \, \mathcal{C}^{(s)PP/SP}_{\mu e} 
\right\vert < 0.7 \,, 
\en 
\eq 
\label{eq:gAAAV-Lambda2}
& &
\left\vert
\mathcal{C}^{(3)AA/AV}_{\mu e}    
\right\vert < 0.2 \,, \quad 
\left\vert
\mathcal{C}^{(c)AA/AV}_{\mu e} 
\right\vert < 1.1 \times 10^{5} \,, \\
& &
\left\vert
\mathcal{C}^{(0)AA/AV}_{\mu e} 
- 0.05 \, \mathcal{C}^{(s)AA/AV}_{\mu e} 
\right\vert < 2.2 \,, \quad 
\left\vert
\mathcal{C}^{(0)AA/AV}_{\mu e} 
+ 0.06 \, \mathcal{C}^{(s) AA/AV}_{\mu e} 
\right\vert < 5.9 \,,
\en
\eq 
\label{eq:gSSPS-Lambda2}
& &
\left\vert
\mathcal{C}^{(3)SS/PS}_{\mu e} 
\right\vert < 60.6 \,, \quad 
\left\vert
\mathcal{C}^{(0)AA/AV}_{\mu e} 
\right\vert < 41.5 \,. 
\en

\end{document}